\documentclass[aps,prl,showpacs,showkeys,amsmath,amssymb,twocolumn,floatfix,superscriptaddress,nofootinbib]{revtex4-1}

\usepackage[colorlinks=true, linkcolor=blue, citecolor=blue, urlcolor=blue, linktoc=page, bookmarks=false, pdfstartview={FitH}, pdfborder={0 0 0.0 [3 3]}]{hyperref} 
\usepackage[utf8]{inputenc}
\usepackage{xcolor}
\usepackage{amsmath,amssymb}
\usepackage{tensor} 
\usepackage{textgreek}
\usepackage{stmaryrd} 
\usepackage{graphicx}
\usepackage{subfiles}
\newcommand{\notinsubfile}[1]{}
\usepackage{float}
\usepackage{tikz}
\usepackage{siunitx}
\usepackage{bm}
\usepackage{todonotes}
\usepackage{xspace}
\usepackage{multirow}
\usepackage{booktabs}
\usepackage[normalem]{ulem} 

\newcommand{\black}{\textcolor{black}}

\newcommand{\Rnu}{$\overline{\nu}_e$}
\newcommand{\Ufive}{$\mathrm{^{235}U}$}
\newcommand{\Ueight}{$\mathrm{^{238}U}$}
\newcommand{\Punine}{$\mathrm{^{239}Pu}$}
\newcommand{\Puone}{$\mathrm{^{241}Pu}$}

\begin{document}

\title{Comprehensive Measurement of the Reactor Antineutrino Spectrum and Flux at Daya Bay}

\newcommand{\IHEP}{\affiliation{Institute~of~High~Energy~Physics, Beijing}}
\newcommand{\Wisconsin}{\affiliation{University~of~Wisconsin, Madison, Wisconsin 53706}}
\newcommand{\Yale}{\affiliation{Wright~Laboratory and Department~of~Physics, Yale~University, New~Haven, Connecticut 06520}} 
\newcommand{\BNL}{\affiliation{Brookhaven~National~Laboratory, Upton, New York 11973}}
\newcommand{\NTU}{\affiliation{Department of Physics, National~Taiwan~University, Taipei}}
\newcommand{\NUU}{\affiliation{National~United~University, Miao-Li}}
\newcommand{\Dubna}{\affiliation{Joint~Institute~for~Nuclear~Research, Dubna, Moscow~Region}}
\newcommand{\CalTech}{\affiliation{California~Institute~of~Technology, Pasadena, California 91125}}
\newcommand{\CUHK}{\affiliation{Chinese~University~of~Hong~Kong, Hong~Kong}}
\newcommand{\NCTU}{\affiliation{Institute~of~Physics, National~Chiao-Tung~University, Hsinchu}}
\newcommand{\NJU}{\affiliation{Nanjing~University, Nanjing}}
\newcommand{\TsingHua}{\affiliation{Department~of~Engineering~Physics, Tsinghua~University, Beijing}}
\newcommand{\SZU}{\affiliation{Shenzhen~University, Shenzhen}}
\newcommand{\NCEPU}{\affiliation{North~China~Electric~Power~University, Beijing}}
\newcommand{\Siena}{\affiliation{Siena~College, Loudonville, New York  12211}}
\newcommand{\IIT}{\affiliation{Department of Physics, Illinois~Institute~of~Technology, Chicago, Illinois  60616}}
\newcommand{\LBNL}{\affiliation{Lawrence~Berkeley~National~Laboratory, Berkeley, California 94720}}
\newcommand{\UIUC}{\affiliation{Department of Physics, University~of~Illinois~at~Urbana-Champaign, Urbana, Illinois 61801}}
\newcommand{\SJTU}{\affiliation{Department of Physics and Astronomy, Shanghai Jiao Tong University, Shanghai Laboratory for Particle Physics and Cosmology, Shanghai}}
\newcommand{\BNU}{\affiliation{Beijing~Normal~University, Beijing}}
\newcommand{\WM}{\affiliation{College~of~William~and~Mary, Williamsburg, Virginia  23187}}
\newcommand{\Princeton}{\affiliation{Joseph Henry Laboratories, Princeton~University, Princeton, New~Jersey 08544}}
\newcommand{\VirginiaTech}{\affiliation{Center for Neutrino Physics, Virginia~Tech, Blacksburg, Virginia  24061}}
\newcommand{\CIAE}{\affiliation{China~Institute~of~Atomic~Energy, Beijing}}
\newcommand{\SDU}{\affiliation{Shandong~University, Jinan}}
\newcommand{\NanKai}{\affiliation{School of Physics, Nankai~University, Tianjin}}
\newcommand{\UC}{\affiliation{Department of Physics, University~of~Cincinnati, Cincinnati, Ohio 45221}}
\newcommand{\DGUT}{\affiliation{Dongguan~University~of~Technology, Dongguan}}
\newcommand{\XJTU}{\affiliation{Department of Nuclear Science and Technology, School of Energy and Power Engineering, Xi'an Jiaotong University, Xi'an}}
\newcommand{\UCB}{\affiliation{Department of Physics, University~of~California, Berkeley, California  94720}}
\newcommand{\HKU}{\affiliation{Department of Physics, The~University~of~Hong~Kong, Pokfulam, Hong~Kong}}
\newcommand{\Charles}{\affiliation{Charles~University, Faculty~of~Mathematics~and~Physics, Prague}} 
\newcommand{\USTC}{\affiliation{University~of~Science~and~Technology~of~China, Hefei}}
\newcommand{\TempleUniversity}{\affiliation{Department~of~Physics, College~of~Science~and~Technology, Temple~University, Philadelphia, Pennsylvania  19122}}
\newcommand{\CGNPG}{\affiliation{China General Nuclear Power Group, Shenzhen}}
\newcommand{\NUDT}{\affiliation{College of Electronic Science and Engineering, National University of Defense Technology, Changsha}} 
\newcommand{\IowaState}{\affiliation{Iowa~State~University, Ames, Iowa  50011}}
\newcommand{\ZSU}{\affiliation{Sun Yat-Sen (Zhongshan) University, Guangzhou}}
\newcommand{\CQU}{\affiliation{Chongqing University, Chongqing}} 
\newcommand{\BCC}{\altaffiliation[Now at ]{Department of Chemistry and Chemical Technology, Bronx Community College, Bronx, New York  10453}} 

\newcommand{\UCI}{\affiliation{Department of Physics and Astronomy, University of California, Irvine, California 92697}} 
\newcommand{\GXU}{\affiliation{Guangxi University, No.100 Daxue East Road, Nanning}} 
\newcommand{\HKUST}{\affiliation{The Hong Kong University of Science and Technology, Clear Water Bay, Hong Kong}} 
\newcommand{\Rochester}{\altaffiliation[Now at ]{Department of Physics and Astronomy, University of Rochester, Rochester, New York 14627}} 

\newcommand{\LSU}{\altaffiliation[Now at ]{Department of Physics and Astronomy, Louisiana State University, Baton Rouge, LA 70803}} 

\newcommand{\NCSF}{\affiliation{New Cornerstone Science Laboratory, Institute of High Energy Physics, Beijing}} 
\author{F.~P.~An}\ZSU
\author{W.~D.~Bai}\ZSU
\author{A.~B.~Balantekin}\Wisconsin
\author{M.~Bishai}\BNL
\author{S.~Blyth}\NTU
\author{G.~F.~Cao}\IHEP
\author{J.~Cao}\IHEP\NCSF
\author{J.~F.~Chang}\IHEP
\author{Y.~Chang}\NUU
\author{H.~S.~Chen}\IHEP
% ...

\collaboration{Daya Bay Collaboration}\noaffiliation

\begin{abstract}
This Letter reports the precise measurement of the reactor antineutrino spectrum and flux based on the full dataset of 4.7 million inverse-beta-decay (IBD) candidates collected at Daya Bay near detectors.
Expressed in terms of the IBD yield per fission, the antineutrino spectra from all reactor fissile isotopes and the specific \Ufive\ and \Punine\ isotopes are measured with 1.3$\%$, 3$\%$ and 8$\%$ uncertainties, respectively, near the 3 MeV spectrum peak in reconstructed energy, reaching the best precision in the world.
The total antineutrino flux and isotopic \Ufive\ and \Punine\ fluxes are precisely measured to be $5.84\pm0.07$, $6.16\pm0.12$ and $4.16\pm0.21$ in units of $10^{-43} \mathrm{cm^2/fission}$.
These measurements are compared with the Huber-Mueller (HM) model, the reevaluated conversion model based on the Kurchatov Institute (KI) measurement and the latest Summation Model (SM2023).
The Daya Bay flux shows good consistency with the KI and SM2023 models but disagrees with the HM model.
The Daya Bay spectrum, however, disagrees with all model predictions.
\end{abstract}

\maketitle


Nuclear reactors produce an essentially pure flux of electron antineutrinos (\Rnu) via $\beta$-decay processes of fission isotopes.
They have been one of the most powerful tools to study neutrino properties from neutrino discovery~\cite{Cowan_1956} to neutrino oscillations~\cite{KamLAND_2002,DayaBay2012,DoubleChooz2011,RENO2012}, and are expected to continue making significant contributions in the upcoming precision era~\cite{JUNO_YB, JUNO2021, JUNOprecision, JUNO_NMO, Sub13_IHEP, Sub13_SuperChooz}.
Accurate knowledge of the energy spectrum and flux of reactor \Rnu\ is important for precision oscillation measurements.

Despite many advances in understanding the reactor \Rnu\ spectrum and flux from both experimental measurements~\cite{Gosgen1986, Rovno1991, Bugey4, SRP1996, Krasn1999, Palo_Verde2001, CHOOZ2002,  Nucifer2016, DayaBay_flux_spec_2015, DayaBay_PRL_evolution2017, DayaBay_PRL_extraction, DayaBay_PRL_evolution2023, DoubleChooz_NaturePhys, RENO_flux_spec_2021, PROSPECT_final, Stereo_Nature, DANSS_2023, DANSS_Neutrino2024, NEOSII_Neutrino2024}
and theoretical calculations, including conversion models~\cite{model_Huber, model_Mueller, Exp_U238, model_KI} and summation models~\cite{model_SM2018, model_SM2023_PRC, model_SM2023_PRL},
there remain anomalies to be resolved.
The measured reactor \Rnu\ flux rate has an overall 6\% deficit with respect to the Huber-Mueller (HM) model~\cite{RAA,RAA_review2022,RAA_review2024}.
This rate anomaly tends to vanish when confronted with the recent measurement at the Kurchatov Institute (KI)~\cite{model_KI} which claims an overestimation of the \Ufive\ contribution in the HM model as first indicated by Daya Bay in Ref.~\cite{DayaBay_PRL_evolution2017}.
The measured reactor \Rnu\ spectrum shape exhibits typically an excess around 5 MeV
not found in both conversion and summation model predictions.
The latest summation model~\cite{model_SM2023_PRC}, noted as SM2023, has been improved with refined $\beta$-decay formalism and recent evaluated nuclear decay data compared to the previous SM2018 model~\cite{model_SM2018}, and gives the complete error budget of the summation method for the first time.
An alternative summation model, noted as SM2023*, incorporates a single empirical parameter $\alpha$ in the $\beta$-transition model that corrects for the pandemonium effect and missing transitions~\cite{model_SM2023_PRC}; and it shows good agreement with the STEREO \Rnu\ spectrum with $\alpha=0.7$~\cite{Stereo_Nature}.
These recent developments provide insight into the possible origin of the anomalies in spectrum and flux.
However, definitive conclusions remain elusive due to limited precision.
More precise measurements of reactor \Rnu\ spectrum and flux would provide further understanding of these discrepancies.
Such measurements also offer accurate data-driven inputs for other experiments aiming at high-precision neutrino oscillation studies such as JUNO~\cite{JUNO_YB, JUNO2021, JUNO_NMO, JUNOprecision}, and a benchmark for future elastic neutrino-nucleus scattering experiments at low-enriched uranium (LEU) reactors~\cite{Snowmass_Rnu}.
Moreover, such measurements offer valuable inputs for sterile neutrino searches, as well as applications in nuclear science and reactor safety~\cite{Snowmass_Rnu, NuclearSafety}.

This Letter reports a comprehensive measurement of reactor \Rnu\ spectrum and flux at Daya Bay, using data collected over the full experimental operation period.
The Daya Bay experiment, consisting of 8 antineutrino detectors (ADs) deployed in 2 near sites (2 ADs each) and 1 far site (4 ADs), detects \Rnu\ from 6 commercial reactor cores~\cite{DayaBay_CPC_2016}.
In commercial LEU reactors, four parent isotopes, i.e. \Ufive, \Ueight, \Punine\ and \Puone, contribute more than 99.7$\%$ of the \Rnu\ flux.
Utilizing the gadolinium doped liquid scintillator technology, Daya Bay detects \Rnu\ via the inverse-beta-decay (IBD) reaction, i.e., $\overline{\nu}_e + p \rightarrow e^+ + n$, with a prompt signal of positron and a delayed signal of neutron captured primarily on Gd (n-Gd).
The experiment operated for 3158 days from 2011 to 2020 and accumulated about 4.7 million n-Gd IBD candidates with its near detectors.
The total reactor \Rnu\ spectrum and flux are measured in terms of the IBD yield aggregating all isotope contributions.
The total spectrum and flux are then decomposed into \Ufive\ and \Punine\ isotopic contributions using the fuel evolution analysis technique developed by Daya Bay~\cite{DayaBay_PRL_evolution2017, DayaBay_PRL_extraction}.
To examine the data to model consistency, the Daya Bay spectra and fluxes are compared with predictions based on the most representative models to date, including HM, KI, and SM2023.
Furthermore, the spectra are unfolded from reconstructed energy to neutrino energy, where the unfolding technique is applied for multiple spectra together, for the first time, by considering their correlation.




The reactor \Rnu\ IBD yield, denoted as $\sigma_f$, can be understood as the number of \Rnu\ per fission multiplied by the IBD cross section~\cite{IBD_Xsection_Vogel}.
It is experimentally defined at Daya Bay for each AD as
\begin{eqnarray}
\sigma_{f} = \dfrac{N_{\rm{IBD}}}{\epsilon_{\rm{IBD}}\ N_{p}\ T_{\rm{eff}} \sum_{r=1}^{6} \dfrac{W_{\rm{th},r}}{4\pi L^2_{r} \overline{E} }}.
\label{eqIBDyield}
\end{eqnarray}
The ${N_{\rm{IBD}}}$ stands for the number of IBD events obtained based on IBD candidates after the nonequilibrium and spent nuclear fuel corrections, background subtraction, as well as oscillation correction.
The denominator of Eq.(\ref{eqIBDyield}) represents the effective fission number, in which
$\epsilon_{\rm{IBD}}$ is the IBD detection efficiency,
${N}_{p}$ is the number of target protons,
${T}_{\rm{eff}}$ is the effective livetime,
${W}_{\rm{th},r}$ is the reactor thermal power,
${L}_{r}$ is the distance between reactor and detector,
$\overline{E} = \sum_{i}^{} f_{i,r} e_{i}$ is the average energy release per fission with
${f}_{i,r}$ the fission fraction
and ${e}_{i}$ the energy release per fission.
The subscript indices $i$ and $r$ stand for the isotope and reactor core respectively.

The systematic uncertainty in the IBD yield measurement arises from the detector, reactor and background effects, with the detector effects being the dominant factor.
The detector uncertainty in the flux rate is 1.2$\%$, incorporating contributions from the proton number, detection efficiency and IBD cross section~\cite{DayaBay_PRD_flux2019}.
The uncertainties of the energy scale~\cite{DYB_PRL_osc_2018}, energy nonlinearity~\cite{DYB_nonlinearity} and inner acrylic vessel effect~\cite{DayaBay_CPC_2016} are further taken into account in the spectrum shape measurement.
The uncertainty of the reactor power is 0.5$\%$ treated as reactor-uncorrelated~\cite{DayaBay_CPC_2016}.
The fission fraction is calculated with the APOLLO2~\cite{Apollo1988} simulation tool by the reactor company and validated by the Daya Bay Collaboration by using the DRAGON~\cite{Dragon} simulation tool.
The fission fraction uncertainty is conservatively set to be 5$\%$ for each isotope to encompass potential inaccuracies arising from simulation tools,
and it is treated as reactor-uncorrelated ~\cite{DayaBay_CPC_2016, An_thesis, Ma_FF_2014, Ma_FF_2017, Apollo2010, Djurcic:2008ny}.
The energy release per fission and its uncertainty are taken from Ref.~\cite{PRC_fission_energy}.
The uncertainties associated with nonequilibrium and spent nuclear fuel effects are both 30$\%$~\cite{DayaBay_CPC_2016, DYB_PRL_osc_2018}.
The uncertainties related to background and oscillation parameters are taken from Ref.~\cite{DayaBay_final_theta13}.

\begin{figure}[t!]
    \hglue -0.2cm
	\centering
	\includegraphics[scale=0.5]{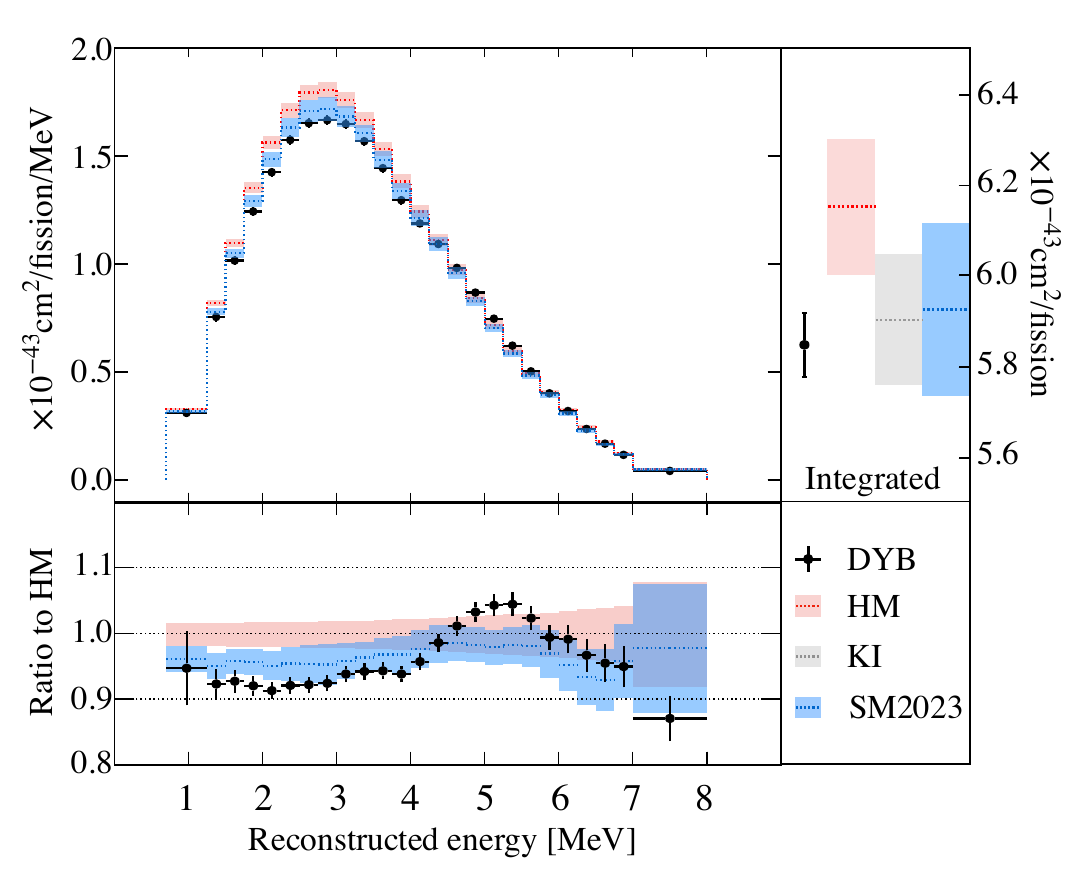}
	\caption
	{
	The total IBD yield spectrum measured at Daya Bay is shown as the black points in the top left panel, in comparison with HM (red)~\cite{model_Huber, model_Mueller} and SM2023 (blue)~\cite{model_SM2023_PRC} models.
	The ratio between data and the HM model \black{is shown in the bottom left panel}, as well as the ratio of SM2023 to HM.
	The top right panel compares the total flux measured by Daya Bay to the HM ($6.15\pm0.15$), SM2023 ($5.92\pm0.19$) and KI~\cite{model_KI} ($5.90\pm0.14$) in units of $10^{-43} \mathrm{cm^2/fission}$.
    The KI spectral shape is consistent with the HM model and omitted from the lower panel for clarity.
	The error bars in the data points represent the square root of the diagonal elements of the covariance matrix for the total spectrum, incorporating both statistic and systematic uncertainties.
	The error bands for different models reflect the uncertainties inherent to each specific model.}
	\label{Fig1}
\end{figure}

\begin{figure*}[]
	\centering
	\includegraphics[scale=0.55]{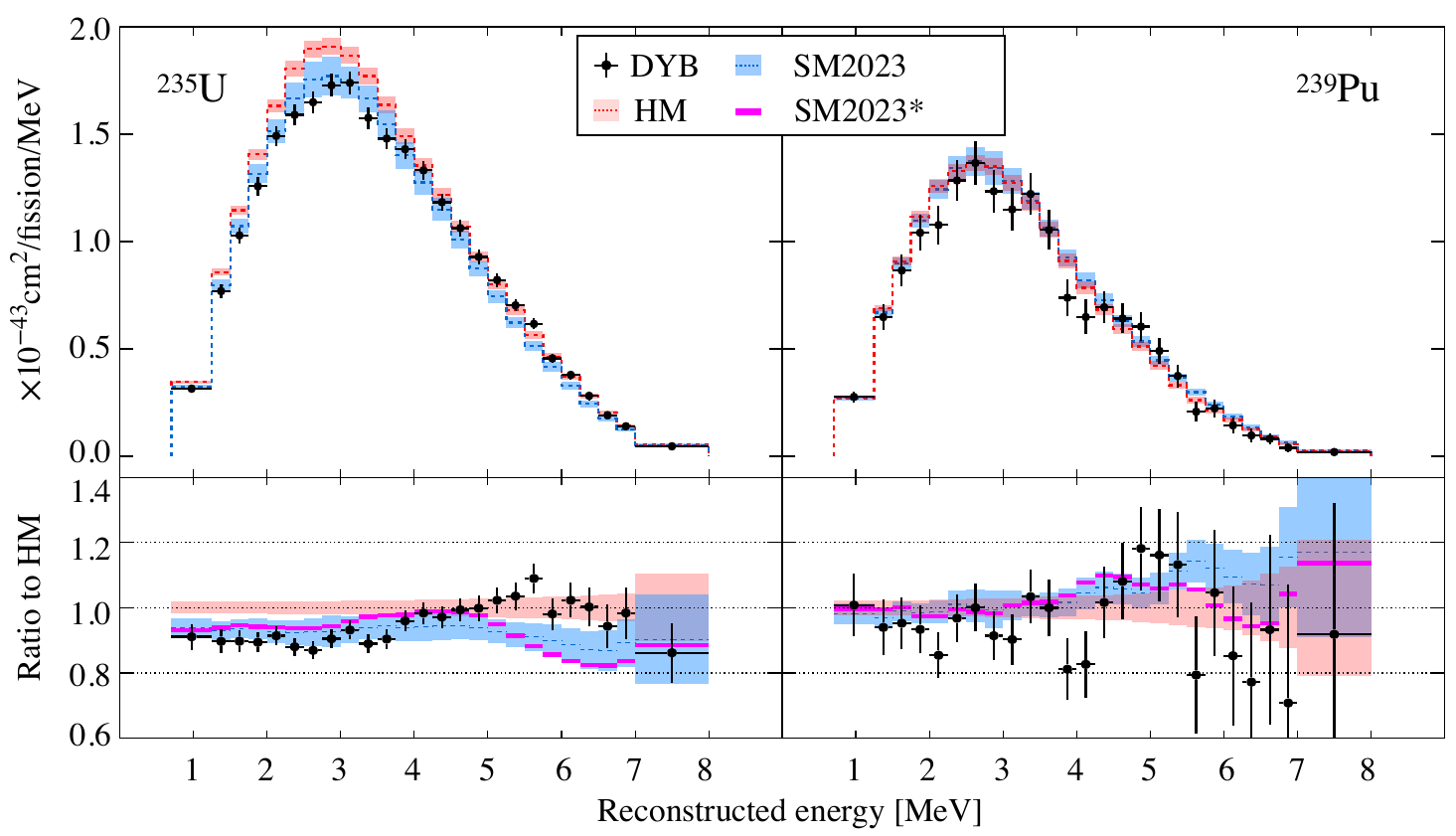}
	\caption
{
	The extracted \Ufive\ (left) and \Punine\ (right) IBD yield spectra in terms of reconstructed energy of prompt signals \black{are shown} as black points.
	For comparison, the HM and SM2023 predictions are drawn in red and blue respectively.
	The SM2023* with the empirical parameter $\alpha = 0.7$ is drawn in magenta.
    The ratios between data and the HM model are shown in the bottom panels, as well as the ratios of SM2023 and SM2023* to HM.
	The error bars in the data points represent the square root of the diagonal elements of the covariance matrix for the extracted spectra.
	The error bands for the HM and SM2023 predictions reflect the uncertainties inherent to each model.
	Only the central values of SM2023* are plotted, as its complete error budget is not provided.
       }
	\label{Fig2}
\end{figure*}

In this analysis, the IBD events are classified into 25 bins according to the reconstructed energy of IBD prompt signals from 0.7 to 8 MeV: 1 bin in 0.7-1.25 MeV, 23 equal bins in 1.25-7 MeV and 1 bin in 7-8 MeV.
According to the definition of Eq.(\ref{eqIBDyield}), Daya Bay measures the total IBD yield spectrum $\bm{s}_f$ using the data of all near ADs, as shown in Fig.~\ref{Fig1}.
Its precision reaches about 1.3$\%$ at the typical reconstructed peak energy around 3 MeV, where the systematic uncertainty is dominant.

In comparison with the HM model, the measured total spectrum shows a disagreement of more than 5$\sigma$ significance with most notably a 8$\%$ deficit below 4 MeV in terms of reconstructed energy.
\black{The SM2023 prediction matches the Daya Bay spectrum better overall at 1.2$\sigma$ significance but shows a worse agreement around the 5 MeV region.}

By integrating all energy bins, the total IBD yield $\sigma_f$ is measured to be $[5.84 \pm0.07] \times 10^{-43} \mathrm{cm^2/fission}$, where the systematic uncertainty is dominant and the statistic uncertainty contributes less than 5$\%$ of the total uncertainty budget.
As shown in the right panel of Fig.~\ref{Fig1}, the Daya Bay flux shows 5.0$\%$ deficit with respect to the HM model, while it is consistent with the KI and SM2023 models.

By factorizing out the rate difference from the models to Daya Bay data, a shape-only comparison is therefore performed.
And it turns out that the Daya Bay spectrum shape shows a clear excess around 5 MeV with respect to the HM model at more than 5$\sigma$ significance and the SM2023 model at 3$\sigma$ significance in the 4 to 6 MeV energy range using the method from Ref.~\cite{DayaBay_CPC_2016}.


The total IBD yield can be considered as the combination of the four major isotopic yields from \Ufive, \Ueight, \Punine\ and \Puone .
Each nuclear isotope yields unique \Rnu\ spectrum and flux due to different fission yields and beta decay branches.
With the burning of reactor fuel, the fraction of isotopes evolves, inducing an evolution of the total IBD yield.
The effective fission fraction $F_{i}$ viewed by one AD is defined as
\begin{eqnarray}
    F_{i} =  \sum_{r=1}^{6} \dfrac{W_{\rm{th},r} f_{i,r}}{L_{r}^{2} \sum_{i}^{} f_{i,r} e_{i} } / \sum_{r=1}^{6} \dfrac{W_{\rm{th},r}}{L_{r}^{2} \sum_{i}^{} f_{i,r} e_{i} },
\label{eqEff_Fiss_Frac}
\end{eqnarray}
to study the IBD yield evolution.
In commercial reactors, during one fuel cycle, the \Ufive\ fraction decreases monotonically,  and the \Ueight\ fraction remains approximately constant, while the \Punine\ and \Puone\ fractions both increase monotonically.
Based on the data from all near ADs, the average effective fission fractions of the 4 isotopes are determined at Daya Bay to be $ \overline{F}_{235} : \overline{F}_{238} : \overline{F}_{239} : \overline{F}_{241} = \black{0.564 : 0.076 : 0.304 : 0.056} $.

The ${F}_{239}$ is chosen to represent the fuel evolution status.
In the following analysis, the ${F}_{239}$ is first calculated on a weekly basis;
then 20 groups are defined based on the weekly ${F}_{239}$ values;
the IBD data is then categorized into the 20 groups, leading to an evolved IBD yield with ${F}_{239}$.
The evolution of the IBD yield rate can be approximated with a linear relation with respect to $F_{239}$, where the slope ${d\sigma_{f}}/{dF_{239}}$ is determined to be $[-1.96\pm0.11\mathrm{(stat.)}\pm0.07\mathrm{(syst.)}] \times 10^{-43} \mathrm{cm^2/fission}$.

The evolution of the total IBD yield spectrum $\bm{s}_f$ enables a decomposition of the isotopic spectra, denoted as $\bm{s}_i$, according to the following $\chi^2$ analysis:
\begin{eqnarray}
\chi^2 = \chi^2(\bm{s}_f, {\bf F}, \bm{s}_i, \bm{\epsilon})  + \chi^2(\bm{s}_{238}, \bm{s}_{241}).
\label{eqSpecDecomp}
\end{eqnarray}
The $\chi^2$ analysis constructs for each $F_{239}$ group the difference between measured total spectrum ($\bm{s}_f$) and corresponding prediction that is the combination of isotopic spectra ($\bm{s}_i$) according to the effective fission fractions $\bf{F}$.
The $\bm{\epsilon}$ represents nuisance parameters encompassing systematic uncertainties from the reactor, detector and background~\cite{DayaBay_CPC_2016, DayaBay_PRD_flux2019, DayaBay_final_theta13}.
The evolution of isotopic fission fractions is degenerate, because the \Punine\ and \Puone\ fractions evolve in a similar manner, and the \Ueight\ fraction is stable during the fuel burning.
In order to reduce the degeneracy and extract the \Ufive\ and \Punine\ spectra, external constraints based on the HM model are introduced on the \Ueight\ and \Puone\ spectra through $\chi^2(\bm{s}_{238}, \bm{s}_{241})$, given that \Ueight\ and \Puone\ are minor contributions at Daya Bay.
The external constraints are loosely set by considering enlarged uncertainties with respect to the HM original ones.
The shape uncertainty of the \Ueight\ spectrum is set to be 10-35$\%$ in 0.7-8 MeV, and the rate uncertainty is set to be $10\%$, which covers the uncertainties from data and also the difference between data and model~\cite{Exp_U238}.
The uncertainty of the \Puone\ spectrum is set to be 7-35$\%$ for the shape, and $10\%$ for the rate.
Consistent results are obtained when the SM2023 model replaces the HM model in the $\chi^2(\bm{s}_{238}, \bm{s}_{241})$ term.

The extracted \Ufive\ and \Punine\ spectra, i.e., $\bm{s}_{235}$ and $\bm{s}_{239}$, are shown in Fig.\ref{Fig2}.
\black{The \Ufive\ and \Punine\ spectra reach an unprecedented precision of 3\% and 8\%, respectively, in the 3 MeV region,} leading to a 15$\%$ improvement compared to previous Daya Bay results~\cite{DayaBay_PRL_extraction}.
The statistical uncertainty still contributes more than 50\% for both spectra.

The \Ufive\ spectrum measured at Daya Bay differs from the HM model \black{with} a deficit below 4 MeV with more than 4$\sigma$ significance.
However, it differs from the SM2023 model \black{most notably with} an excess between 5 and 7 MeV which reaches about 3$\sigma$ significance.
The SM2023* model, despite showing agreement with the STEREO \Ufive\ spectrum~\cite{Stereo_Nature}, disagrees with the Daya Bay \Ufive\ spectrum above 5 MeV as illustrated in Fig.2.
The precision of the Daya Bay \Punine\ spectrum is insufficient to differentiate between models.

In terms of the shape-only comparison, the Daya Bay \Ufive\ and \Punine\ spectra differ from both the HM and SM2023 models with a bump around 5 MeV.
Based on the method detailed in Ref.~\cite{DayaBay_CPC_2016, DayaBay_PRL_extraction}, the local shape-only discrepancy from 4 to 6 MeV for the \Ufive\ spectrum is quantified at 4$\sigma$ level when comparing the Daya Bay measurement with both the HM and SM2023 models, while that for the \Punine\ spectrum is less than 2$\sigma$ due to the relatively large uncertainty in the measurement.

The fission fractions of \Punine\ and \Puone\ exhibit an approximate proportionality.
Thus, as proposed in Ref.~\cite{DayaBay_PRL_extraction}, the reconstructed energy spectra of \Punine\ and \Puone\, can be treated as a single combined component, named as Pu combo.
This is defined as $\bm{s}_{\rm{combo}} = \bm{s}_{239} + k \bm{s}_{241}$, where the $\bm{s}_{\rm{combo}}$ stands for the spectrum of Pu combo and the coefficient $k = 0.185$ is derived by fitting the correlation between $F_{239}$ and $F_{241}$.
The Pu combo approach reduces the reliance on the model for $\bm{s}_{241}$ and the relative uncertainty in $\bm{s}_{\rm{combo}}$ is 30\% less than that of $\bm{s}_{239}$ with negligible impact on $\bm{s}_{235}$.
The Supplemental Material contains the $\bm{s}_{\rm{combo}}$ results.

\begin{figure}[t!]
	\centering
	\hglue -0.5cm
	\includegraphics[width=0.9\linewidth]{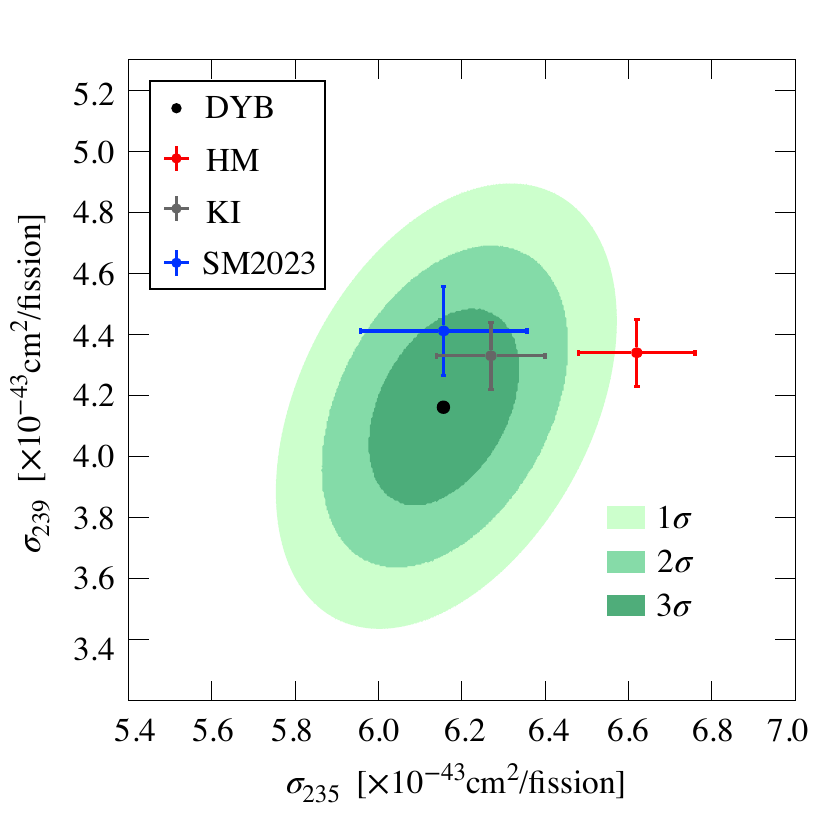}
	\caption{
	The extracted isotopic reactor \Rnu\ fluxes of \Ufive\ and \Punine\ in terms of IBD yield are represented by the black point.
	The green contours indicate the $\sigma_{235}$ and $\sigma_{239}$ two-dimensional allowed regions with 1$\sigma$, 2$\sigma$, and 3$\sigma$ significance.
    For comparison, the HM, KI and SM2023 model values are provided for $\sigma_{235}$ as $6.62 \pm 0.14$, $6.27 \pm 0.13$ and $6.16 \pm 0.20$, and for $\sigma_{239}$ as $4.34 \pm 0.11$, $4.33 \pm 0.11$ and $4.41 \pm 0.15$, in units of $10^{-43} \mathrm{cm^2/fission}$.
    }
	\label{Fig3}
\end{figure}

In analogy with the spectrum decomposition, the \Ufive\ and \Punine\ flux rates, denoted as $\sigma_{235}$ and $\sigma_{239}$, can also be extracted through a $\chi^2$ analysis similar to~Eq.(\ref{eqSpecDecomp}) by replacing 25 energy bins by an integrated bin.
The rate decomposition gives $\sigma_{235}$ and $\sigma_{239}$, respectively, as
$[6.16 \pm 0.04 \mathrm{(stat.) \pm 0.08 \mathrm{(syst.)} \pm 0.08 \mathrm{(model)}}]$
and
$[4.16 \pm 0.07 \mathrm{(stat.) \pm 0.09 \mathrm{(syst.)} \pm 0.18 \mathrm{(model)}}] \times 10^{-43} \mathrm{cm^2/fission}$, as shown in Fig.\ref{Fig3}.
The systematic uncertainty and the uncertainty from the model constraint of \Ueight\ and \Puone\ are equally dominant for $\sigma_{235}$, while the uncertainty from the model constraint is dominant for $\sigma_{239}$.
The precision of $\sigma_{235}$ and $\sigma_{239}$ has been improved by about 15\% compared to the previous results in Ref.~\cite{DayaBay_PRL_extraction}.
Daya Bay determines the ratio $\sigma_{235}/\sigma_{239} = 1.48\pm0.07$ consistent with the measurements from DANSS~\cite{DANSS_2023}, NEOS-II~\cite{NEOSII_Neutrino2024} and RENO~\cite{RENO_talk_Neutrino2022}.

The Daya Bay \Ufive\ flux shows a $7.0\%$ deficit in comparison with the HM value with about 3$\sigma$ significance,
\black{while it is only $1.8\%$ lower than the KI prediction and
well consistent with the SM2023 prediction.}
\black{The Daya Bay \Punine\ flux shows respective $4.1\%$, $3.9\%$ and $5.7\%$ deficit to the HM, KI and SM2023 model predictions in this analysis with full dataset, while it was consistent with models in previous measurements using partial data~\cite{DayaBay_PRL_evolution2017, DayaBay_PRL_extraction}.}
However, considering its relatively large uncertainty, this deficit has only about 1$\sigma$ significance.

Thanks to the flux decomposition analysis, Daya Bay is able to examine the possible contributing isotopes responsible for the reactor antineutrino anomaly in total rate with respect to the HM model. Hence, several hypothesis tests are performed.
In the previous Daya Bay publication~\cite{DayaBay_PRL_evolution2017}, only the HM central values are considered in the hypothesis test, whereas a more inclusive approach is adopted here by also taking into account the uncertainties of the HM model.
Compared to the case where both \Ufive\ and \Punine\ contribute to the deficit in total rate,
the hypothesis of \Ufive\ as the sole contributor is slightly disfavored by 0.9$\sigma$;
the hypothesis of \Punine\ as the sole contributor is more significantly disfavored by 2.6$\sigma$;
moreover, the hypothesis of the four isotopes as the contributors is merely disfavored by 1.4$\sigma$.
Because of the deficits observed in both \Ufive\ and \Punine\ in the final dataset, Daya Bay has no strong preference for \Ufive\ as the sole offending isotope in the HM prediction.




The aforementioned measurements on total, \Ufive\ or \Punine\ IBD yield spectra are achieved in terms of reconstructed energy for positron signals, which contains detector response effects, such as energy scale nonlinearity and resolution.
The neutrino energy spectrum can be obtained by applying unfolding technique, such as singular value decomposition (SVD) regularization~\cite{method_SVD}, Wiener SVD~\cite{method_WSVD} and Bayesian iteration~\cite{method_Bayesian} methods.
The unfolded neutrino energy spectrum facilitates a direct comparison with models or experiments, and can also serve as an input spectrum for other reactor neutrino experiments.

As Daya Bay extracts simultaneously the reconstructed energy spectra of \Ufive\ and \Punine\ , where correlation exists between $\bm{s}_{235}$ and $\bm{s}_{239}$, the correlation should be taken into account when performing unfolding.
In addition, as shown in Ref.~\cite{DYB_unfolding}, a generic unfolded neutrino energy spectrum ($\bm{s}^{\nu}_g$) can be constructed primarily based on the Daya Bay's total spectrum($\bm{s}^{\nu}_f$), with isotopic spectra ($\bm{s}^{\nu}_{i}$) as corrections according to the fission fraction difference between experiments.
In this context, the correlation among the Daya Bay reconstructed energy spectra of $\bm{s}_{f}$, $\bm{s}_{235}$ and $\bm{s}_{239}$ should all be included when performing unfolding.

\begin{figure}
	\hglue -0.3cm
	\centering
	\includegraphics[scale=0.45]{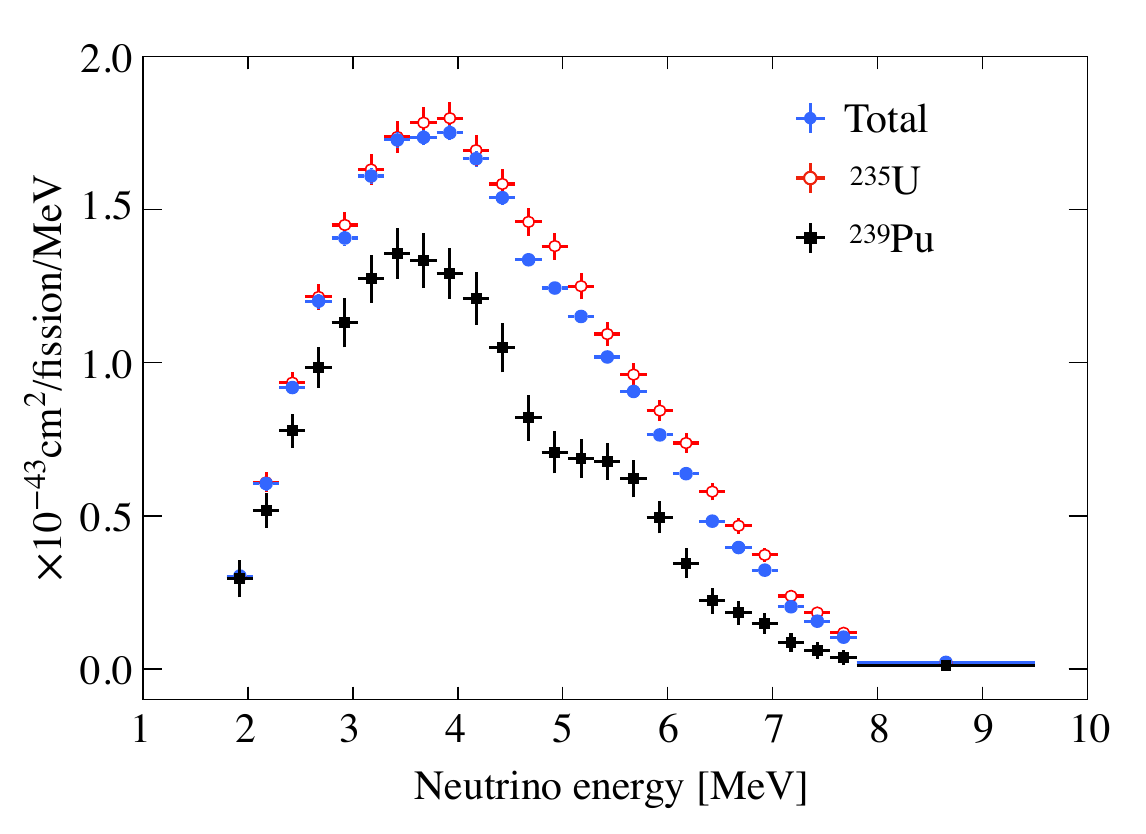}
	\caption
	{
	Unfolded neutrino energy spectra for the total, \Ufive\ and \Punine\ IBD yields.
	The unfolding imposes a smoothness condition while the characteristic spectral features persist in the energy range between 5 and 7 MeV.
	The error bars for the data points represent the square root of the diagonal elements of the covariance matrix.
    }
	\label{Fig_unfolding}
\end{figure}

The Wiener-SVD method achieved a smaller mean squared error (MSE) than other methods in the previous Daya Bay analysis where each spectrum was unfolded individually~\cite{DYB_unfolding}.
The correlation among spectra affects the Wiener filter undesirably but does not affect the traditional SVD regularization.
Therefore, the SVD regularization method is adopted to unfold the $\bm{s}_{f}$, $\bm{s}_{235}$ and $\bm{s}_{239}$ together, which is achieved by minimizing the following $\chi^2$:
\begin{eqnarray}
    \chi^2 &=&
    \left( \bm{S} - \bm{R} \bm{S}^{\nu} \right)^T\bm{V}^{-1}\left( \bm{S} - \bm{R} \bm{S}^{\nu} \right) \nonumber\\
    & + & \tau  \left(\bm{C}\bm{S}^{\nu} \right)^T\left(\bm{C}\bm{S}^{\nu} \right).
\end{eqnarray}
$\bm{S}$ is composed of the three reconstructed energy spectra,  $\bm{s}_{f}$, $\bm{s}_{235}$ and $\bm{s}_{239}$,
and $\bm{S}^{\nu}$ is composed of the corresponding neutrino energy spectra, $\bm{s}^{\nu}_{f}$, $\bm{s}^{\nu}_{235}$ and $\bm{s}^{\nu}_{239}$.
$\bm{V}$ is the covariance matrix of the three reconstructed energy spectra.
$\bm{R}$ is the response matrix, which \black{contains} the conversion relation between reconstructed energy and neutrino energy, and it contains three submatrices for the three spectra.
$\tau$ is the regularization strength that minimizes the MSE between data and the model.
$\bm{C}$ is composed of three second-order derivative matrices arranged diagonally, through which the smoothness of each individual spectrum is imposed.
The total, \Ufive\ and \Punine\ spectra are unfolded together, bringing about the neutrino energy spectra as presented in Fig.~\ref{Fig_unfolding}.
Consistent results are obtained when using different models for the MSE minimization.
The additional smearing matrices and covariance matrices associated with the unfolded neutrino energy spectra, as well as the unfolding inputs can be found in Supplemental Material.


In summary, this Letter presents a comprehensive measurement of reactor \Rnu\ spectrum and flux based on the full IBD dataset collected with Daya Bay near detectors.
The reactor antineutrino spectra and fluxes from \Ufive\ and \Punine\ are extracted using the fuel evolution technique.
In terms of the reconstructed energy of prompt signals, the total, \Ufive\ and \Punine\ spectra are measured with precision 1.3$\%$, 3$\%$ and 8$\%$, respectively, around the 3 MeV peak energy.
The Daya Bay spectra are inconsistent with model predictions, exhibiting most notably a 5 MeV bump in shape.
When considering the comparison in both shape and rate, the Daya Bay \Ufive\ spectrum shows a significant deficit below 4 MeV compared to the HM prediction.
In addition, Daya Bay measures the energy-integrated total, \Ufive\ and \Punine\ fluxes with 1.2$\%$, 1.9$\%$ and 5.0$\%$ precision, respectively.
The total and \Ufive\ fluxes of Daya Bay show, respectively, 5.0$\%$ and 7.0$\%$ deficits compared to the HM predictions at about 3$\sigma$ significance level; however, they are consistent with the KI and SM2023 models.
Moreover, the measured three reconstructed energy spectra are unfolded into neutrino energy spectra using the SVD regularization method with correlations among different spectra taken into account when simultaneously unfolding multiple spectra together.
The world-leading precision measurement presented in this Letter enriches the knowledge of the reactor \Rnu\ spectrum and flux.
The discrepancies presented in this Letter reinforce the necessity of higher precision measurements and the further refinement of models in the future. \\

The Daya Bay experiment is supported in part by the Ministry of Science and Technology of China, the U.S. Department of Energy, the Chinese Academy of Sciences, the CAS Center for Excellence in Particle Physics, the National Natural Science Foundation of China, the New Cornerstone Science Foundation, the Guangdong provincial government, the Shenzhen municipal government, the China General Nuclear Power Group, the Research Grants Council of the Hong Kong Special Administrative Region of China, the Ministry of Education in Taiwan, the U.S. National Science Foundation, the Ministry of Education, Youth, and Sports of the Czech Republic, the Charles University Research Centre UNCE, and the Joint Institute of Nuclear Research in Dubna, Russia. We acknowledge Yellow River Engineering Consulting Co., Ltd., and China Railway 15th Bureau Group Co., Ltd., for building the underground laboratory. We are grateful for the cooperation from the China Guangdong Nuclear Power Group and China Light $\&$ Power Company.

\nocite{*}
\bibliography{ref}

\end{document}